\documentclass[apsprb,twocolumn,superscriptaddress,amsmath,amssymb,showpacs]{revtex4-1}
\usepackage{graphicx}
\usepackage{bm}
\usepackage{txfonts}
\usepackage{multirow}

\begin{document}
\title{Sign identification of electron and hole out-of-plane {\rm g} factors \\
by utilizing nuclear spin switch in single quantum nanostructures}

\author{R.\ Matsusaki}
	\affiliation{Division of Applied Physics, Graduate School of Engineering, Hokkaido University, N13 W8, Kitaku, Sapporo 060-8628, Japan}
	
\author{R.\ Kaji}
\email{r-kaji@eng.hokudai.ac.jp}
	\affiliation{Division of Applied Physics, Graduate School of Engineering, Hokkaido University, N13 W8, Kitaku, Sapporo 060-8628, Japan}

\author{S.\ Yamamoto}
	\affiliation{Division of Applied Physics, Graduate School of Engineering, Hokkaido University, N13 W8, Kitaku, Sapporo 060-8628, Japan}	
\author{S.\ Adachi}
\email{adachi-s@eng.hokudai.ac.jp}
	\affiliation{Division of Applied Physics, Graduate School of Engineering, Hokkaido University, N13 W8, Kitaku, Sapporo 060-8628, Japan}	

\date{\today}
\begin{abstract}
The electron and hole g factors are the key quantities for the spin manipulations in semiconductor quantum nanostructures. However, for the individual nanostructures, the separate determination including the signs of those g factors is difficult by using some methods adopted conventionally in bulks and quantum wells. We report a convenient optical method for the sign identification of out-of-plane g factors in the individual quantum nanostructures, which utilizes the optically-induced nuclear spin switch. The method is demonstrated in typical single self-assembled In$_{0.75}$Al$_{0.25}$As/Al$_{0.3}$Ga$_{0.7}$As quantum dots and InAs/GaAs quantum rings, where the g factors with the opposite sign for electron and the same sign for hole are proved.
\end{abstract}
\maketitle

The carrier spin dynamics has always been one of the central topics in semiconductor spin physics~\cite{OptOrientation,SpinPhysics}. In particular, the spin manipulations by electrical, magnetic and/or optical methods have been studied intensively~\cite{SpinPhysics,Slavcheva10,Bennett13,Nakaoka07}.
One of the key quantities for the control protocols of localized carrier spins is the g factor, which is the coefficient connecting its magnetic dipole moment with the spin degrees of freedom. In general, the effective g factor in a semiconductor nanostructure deviates from the value dominantly determined by the material composition due to a wide variety of the modulations: for example, the spatial confinement in the nanostructures, the effect of strain-induced valence band mixing (VBM), and the penetration of the carrier wavefunction into the barrier material~\cite{Snelling91,Sirenko97,Nakaoka04,Tholen16}. 

Especially in semiconductor quantum nanostructures such as dots (QDs) and rings (QRs), some of such modulations are enhanced. The electron g factor is determined by a balance between the bare-electron contribution (+2.0) and a lattice orbital contribution, which may vary significantly in both magnitude and sign by strength of the confinement. Also, it has been found that the sign inversion of hole g factor is possible by the external means~\cite{Bennett13}. The sign of g factor is obviously important because it dominates the energy diagram of target spins under the external and/or some effective magnetic fields, and consequently the signs of Zeeman splitting energy and spin precession direction. In order to realize composite spin rotations in a Bloch sphere by using magnetic fields, the sign of g factor determine the rotation direction and thus, the optimum shortest path in a Bloch sphere~\cite{Slavcheva10}. Further, the direction of the photo-induced nuclear field also depends on the sign of g factor as shown later.

In the QD and QR structures, the electron and hole g factors gain another significant importance because the former is important to describe the dynamics of a coupled spin system with nuclei and the latter is essential to measure the VBM. This is because the hyperfine interaction (HFI) and VBM are enhanced for localized carriers due to the strong confinement and large residual strain compared to high dimensional structures like bulks and quantum wells (QWs). Since the effective g factors have the distribution even in the same sample, the individual measurement for target QDs is crucial.

Besides, several works have focused on developing techniques for the evaluation of electron and hole effective g factors (${g}^{\rm e},\ {g}^{\rm h}$), and some methods to determine unambiguously the sign of ${g}^{\rm e}$ have been demonstrated in QWs: the time-resolved photoluminescence (TR-PL) with the up-conversion technique~\cite{Marie00}, TR-Kerr (or Faraday) rotation~\cite{Malinowski00,Salis01}, and Hanle effect measurements~\cite{Snelling91,Malinowski99}. In these methods, the external magnetic field is applied in the sample plane, and the optical axis of incident light is inclined with respect to the sample growth axis. The phase shift of the electron spin precession observed in the TR techniques and the peak shift of Hanle curve caused by an optically-induced nuclear field ($B_{\rm n}$) are key points for the sign determination of ${g}^{\rm e}$. 
However, there are some problems to apply these methods to \textit{individual} QDs and QRs; the TR measurements require relatively large number of electrons and is not successful in single self-assembled (SA) QDs.
Although the Hanle effect measurement is possible for single QDs, the peak shift of Hanle curve is obscured by more prominent deformation ``anomalous Hanle effect" due to the enhanced quadrupolar effect of nuclear spins~\cite{Krebs10,Yamamoto18}. 
Further, if both the bright and dark exciton emissions appear under a longitudinal magnetic field, the electron and hole out-of-plane g factors (${g}^{\rm e}_z$, ${g}^{\rm h}_z$) can be deduced including their signs by using the standard time-integrated micro-PL ($\mu$-PL) measurements. Although this method can be used for the (001)-oriented QDs with reduced symmetry in the shape~\cite{Bayer99,Chekhovich10} and the (111)-oriented QDs~\cite{Sallen11}, the selection of QDs is required. Therefore, the convenient method applicable for the standard single QDs is highly aspired.

In our previous work~\cite{Kaji07}, we demonstrated the method to evaluate $|g_{z}^{\rm e}|$ and $|g_{z}^{\rm h}|$ separately by using the nuclear spin switch (NSSW)~\cite{Tartakovskii07}. Although it was a powerful tool to know the \textit{magnitude} of g factors, we got only the information of the sign of a product $g_{z}^{\rm e} \cdot g_{z}^{\rm h}$.
In this paper, we demonstrate a useful method for separate evaluations of the signs of $g_{z}^{\rm e}$ and $g_{z}^{\rm h}$ as well as their magnitudes. The method utilizes the correlation of the NSSW and the sign of Zeeman splitting in the $\mu$-PL measurements of single QDs.  
The method is demonstrated in typical single SA In$_{0.75}$Al$_{0.25}$As/Al$_{0.3}$Ga$_{0.7}$As QDs and InAs/GaAs QRs, and the observed g factors with the opposite sign for electron and the same sign for hole are proved. 
Since the electron and hole g factors in widely-used In(Ga)As and GaAs QDs have been studied experimentally and theoretically~\cite{Nakaoka04,Bayer99,Sallen11,Kaji14}, the carrier g factors of InAlAs QDs, which have only a few reports, have to be investigated and compared by using an unified method.

We carried out the $\mu$-PL measurements in the time-integrated mode at 6 K under a longitudinal magnetic field $\bm B_{z}$. A continuous wave Ti:sapphire laser was tuned to provide the transition energy to the foot of the wetting layer. The corresponding wavelengths were $\sim$730 nm for InAlAs QDs and $\sim$865 nm for InAs QRs. The excitation beam was focused on the sample surface using a microscope objective lens ($\times$20, NA$\sim$0.4), and the PL signals were collected by the same objective lens and were detected by a triple-grating spectrometer and a liquid N$_{2}$-cooled Si-CCD detector. The spectral resolution that determines the PL energies was $\le$5 $\mu$eV using the spectral fitting. The excitation light polarization was varied systematically by a set of a linear polarizer, a rotating half waveplate, and a quarter waveplate with a fixed angle ($\pi/4$ rad.), and the polarization was monitored by a polarimeter.

\begin{figure}[t]
  \begin{center}
    \includegraphics[width=165pt]{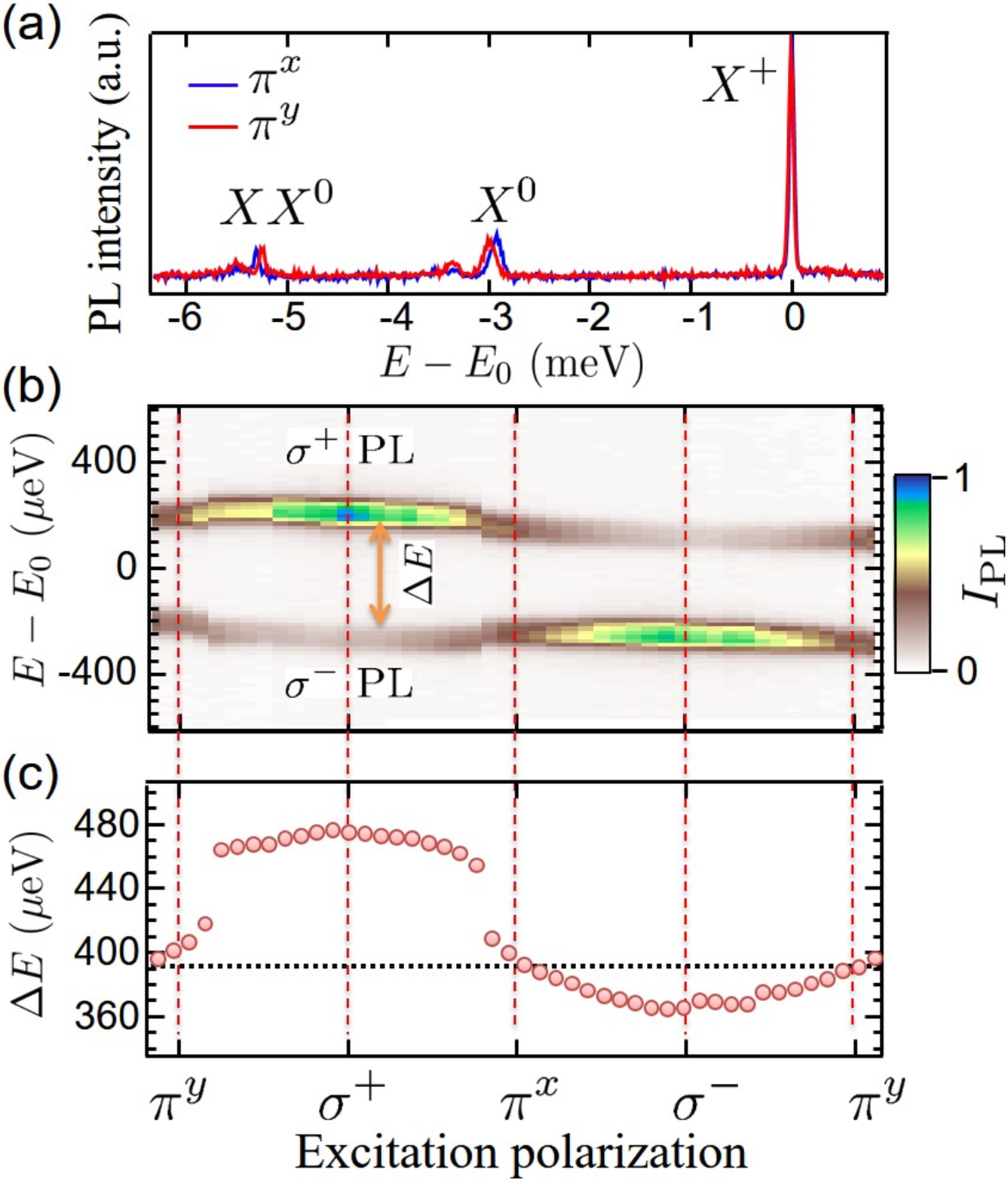}  
  \end{center}
  \caption{(color online) (a) Polarization-resolved  PL spectra of a typical single InAlAs QD at 0 T under the nonpolarized excitation. The horizontal axis is replotted from the $X^{+}$ PL peak energy $E_0$=1.6449 eV. The FSS of $\sim$73 $\mu$eV, the inverse pattern of FSS in the $X^{0}$ and $XX^{0}$ peaks, and no splitting in the $X^+$ peak are observed clearly. (b) Two-dimensional plot of the $X^+$ PL spectra varying the retardance of the excitation polarization systematically at $+3.0$ T. Some light polarizations are indicated in the horizontal axis. The PL spectrum in a high (low) energy side is $\sigma^{+}$ ($\sigma^{-}$) polarized. The red-shift of both PL peaks occurs slightly by the sample heating. (c) The splitting energy between the $\sigma^{+}$ and $\sigma^{-}$ PL peaks ($\Delta E$) as a function of the excitation polarization. The abrupt changes of $\Delta E$ are seen clearly around the $\sigma^{+}$ excitation. The horizontal dotted line indicates the Zeeman splitting energy ($\sim$390 $\mu$eV) by a linearly polarized excitation.}
  \label{Fig1}
\end{figure}

First, we show the results in InAlAs QDs following the sign identification method of the g factors. 
The SA-InAlAs/AlGaAs QDs grown on an undoped (100)-GaAs substrate by molecular beam epitaxy were used. 
The average diameter, height, and density of the QDs were found to be $\sim$20 nm, $\sim$4 nm, and $\sim$5$\times 10^{10}$ cm$^{-2}$, respectively, by the atomic force microscopy measurements of a reference uncapped QD layer~\cite{Yokoi05}. 
It is known that InAlAs/Al(Ga)As QDs have the complex band structure depending on the aluminum concentration and the QD size~\cite{Debus14}. 
In the studied InAlAs QDs, the lowest electron level appears always in the $\Gamma$ valley, and the resultant direct gap structure gives a short recombination time of $\sim$1 ns to the QD exciton~\cite{Kumano06}. 
Figure~\ref{Fig1}(a) shows the polarization-resolved PL spectra ($\pi^{x}$, $\pi^{y}$) of a typical single InAlAs QD at 6 K and 0 T under the nonpolarized excitation. 
The spectra indicate the emissions of the neutral biexciton ($XX^{0}$), neutral exciton ($X^{0}$), and positive trion ($X^{+}$) from the low energy side.  
Each charge state could be assigned by considering the fine structure splitting (FSS) and the binding energy~\cite{Kaji_Thesis}. 
The fact that these PL peaks originate from the same single QD can be confirmed with observing the response to the generated nuclear field by the circularly polarized excitation~\cite{Sasakura08}.

\begin{figure}[t]
  \begin{center}
    \includegraphics[width=220pt]{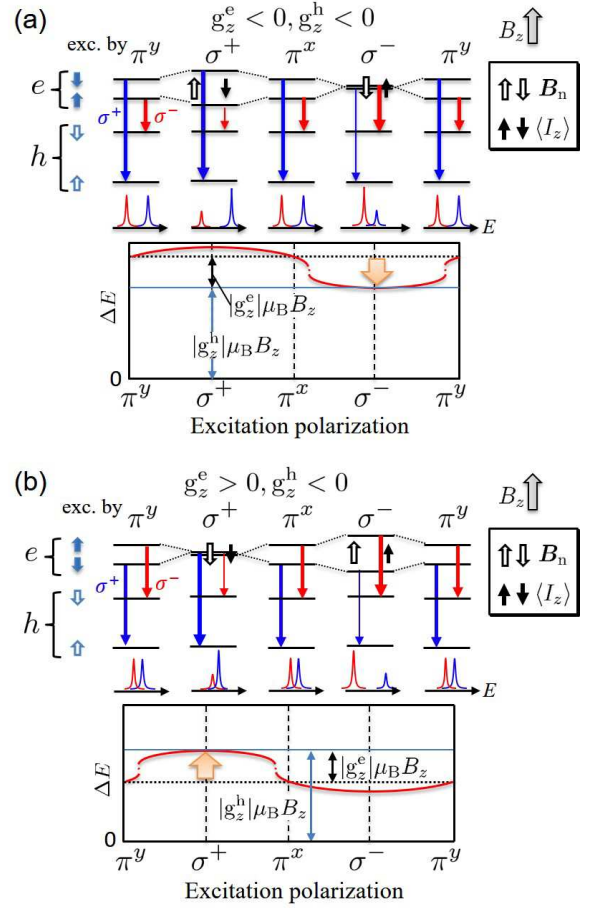}  
  \end{center}
  \caption{(color online) (a) (upper panel) In the case of ${g}_{z}^{\rm e}<0$ and ${g}_{z}^{\rm h}<0$, the diagram of the electron and hole spin states and the expected PL spectra under a longitudinal field $B_{z}$. (lower panel) The splitting energy $\Delta E=E(\sigma^+)-E(\sigma^-)$ as a function of the excitation polarization. The abrupt \textit{decrease} of $\Delta E$ occurs with $\sigma^{-}$. (b) (upper and lower panels) The similar diagrams and the change of $\Delta E$ in the case of ${g}_{z}^{\rm e} > 0$ and ${g}_{z}^{\rm h} < 0$. The abrupt \textit{increase} of $\Delta E$ occurs with $\sigma^{+}$.
Note that the cancellation of $B_z$ by $B_{\rm n}$ occurs by $\sigma^{-}$ excitation in (a) and  by $\sigma^{+}$ excitation in (b).}
  \label{Fig2}
\end{figure}

Figure~\ref{Fig1}(b) shows a two-dimensional plot of the $X^{+}$ PL spectra as a function of the excitation polarization where the retardance of the excitation light is varied systematically under $B_{z}$=$+3.0$ T. The $\sigma^{+}$ polarized PL peak appears at higher energy than the $\sigma^{-}$ polarized one. The intensity and energy of each PL spectrum change clearly depending on the excitation polarization. Figure~\ref{Fig1}(c) indicates the change in energy splitting ($\Delta E$). $\Delta E$ is defined as $\Delta E=E(\sigma^+)-E(\sigma^-)$, where $E(\sigma^{+(-)})$ is the energy of the $\sigma^{+(-)}$ polarized PL peak.
While $\Delta E$ decreases gradually and shows the minimum around the $\sigma^{-}$ excitation, an abrupt increase of $\Delta E$ occurs by NSSW around the $\sigma^{+}$ excitation. 
This change of $\Delta E$ by the generated nuclear field $\bm B_{\rm n}$ (Overhauser field) and the connection to the sign of the g factors will be explained later by using Fig.~\ref{Fig2}. Note that $\Delta E$ stabilizes enough within the exposure time of a CCD detector (1 s) because the formation time of the steady-state $\bm B_{\rm n}$ is $\sim$10 ms at the excitation condition.

Throughout this paper, we consider only the dipole-allowed transitions between the lowest electron states $|S_z \rangle =|\pm 1/2 \rangle$ and the lowest mainly heavy-hole states $|J_z^{\rm h} \rangle =|\pm 3/2 \rangle$. According to the selection rules of optical transitions, the $\sigma^{+}$ polarization triggers the transition $|- 1/2 \rangle (\downarrow) \leftrightarrow |+ 3/2 \rangle (\Uparrow)$, and the $\sigma^{-}$ polarization triggers the transition $|+ 1/2 \rangle (\uparrow) \leftrightarrow |- 3/2 \rangle (\Downarrow)$.
Then, the sign of g factor is defined as follows; the positive (negative) sign of $g_{z}^{\rm e}$ corresponds to the up-spin (down-spin) electron state in the higher energy side compared with the down-spin (up-spin) electron state under the positive $B_{z}$. This definition comes from the standard relation between the magnetic moment $\bm{\mu}$ and the electron spin $\bm{S}$: $\bm{\mu} = -g^{\rm e} \mu_{\rm B} \bm{S}$ ($\mu_{\rm B}$: the Bohr magneton). 
In the case of the hole states, the down-spin (up-spin) hole state is in the higher energy side  under $B_z>0$ according to the relation between the magnetic moment and spin of hole (positive charge), and then the sign is defined as $g_z^{\rm h}<0$ ($g_z^{\rm h}>0$) in this work~\cite{comment}. These definitions of polarization and the signs of $g_z^{\rm e}$ and $g_z^{\rm h}$ are indicated clearly in the state diagrams of Fig.~\ref{Fig2}(a) and (b).

In the demonstrated sign identification method of the g factors, we utilize the NSSW, which means the cancellation of $B_z$ by $B_{\rm n}$. In order to utilize the NSSW, the electron spin interacting with nuclear spins has to be defined; the optically-injected electron spin polarization is highly preserved after the relaxation to the lowest electron level of $X^0$ and $X^+$ in the measurements, which means that the electron and hole spin relaxation times are longer than the radiative recombination time. 
This condition can be realized generally even by the nonresonant excitation of the wetting layer as well as the resonant and quasi-resonant (e.g. 1LO) excitations. As a result, the same polarized PL peak as the excitation polarization has a larger intensity. 
In the case of $X^-$, since the remained electron spin after the e-h recombination interacts with the nuclei, the situation becomes complicated. However, the proposed method can be applied, which will be discussed later. 

In this paper, we set the following conditions in order to keep the representation simple. 
\begin{enumerate}
  \item The absolute value of $|g_{z}^{\rm h}|$ is larger than $|g_{z}^{\rm e}|$, which is adopted generally in III-V QDs, and the condition is satisfied for the investigated InAlAs QDs and InAs QRs. 
  \item The averaged HFI constant $\tilde{A}$ of the QD materials is positive. Since the HFI constants $A$ for all isotopes of III-V elements are positive~\cite{OptOrientation}, $\tilde{A}$ of the popular III-V QDs (GaAs, In(Ga, Al)As, InP) is positive~\cite{Testelin09}. 
  \item The PL spectra from the ground state of $X^+$ or $X^0$ are considered.
\end{enumerate}

The upper panels of Fig.~\ref{Fig2}(a) and (b) are the excitation polarization dependence of the electron and hole energy states under $B_{z}(>0)$ affected by the generated $\bm B_{\rm n}$ in the cases of (a) ${g}_{z}^{\rm e}<0$, ${g}_{z}^{\rm h}<0$ and (b) ${g}_{z}^{\rm e}>0$, ${g}_{z}^{\rm h}<0$. The dipole-allowed transitions with the PL polarization and the expected Zeeman-split spectra (red: $\sigma^-$, blue: $\sigma^+$) are depicted schematically. 

Since the angular momentum of a photon is transferred onto nuclei via the flip-flop term of HFI~\cite{OptOrientation}, a macroscopic nuclear spin polarization (NSP, $\langle {I_z} \rangle$) which is orders of magnitude larger than the value in thermal equilibrium can be generated actually at cryogenic temperatures, and in turn, the resultant $\bm B_{\rm n}$ affects the electron spin states in the Zeeman splitting as well as the dynamics significantly~\cite{Gammon01,Yokoi05,Kaji07,Chekhovich10,Urbaszek13}.
Note that the $\bm B_{\rm n}$ affects only the electron states, not hole states because of the nonzero existence probability of the Bloch function at the nucleus site as shown in the energy diagrams of Fig.~\ref{Fig2}. 
For the linearly polarized excitation ($\pi^x$ and $\pi^y$), the $\sigma^{+}$ and $\sigma^{-}$ polarized PL spectra should indicate in principle the equivalent PL intensity and the $\Delta E$ is determined only by $B_z$ (i.e., $B_{\rm n}$=0 if averaged electron spin polarization $\langle S_z \rangle$=0). In the case of $\langle S_z \rangle \neq$0, the nonzero $B_{\rm n}$ is generated and affects the energy shift of the electron spin states.

The flip-flop term of the collinear HFI builds up the NSP $\langle I_{z} \rangle$ always parallel to the electron spin $S_z$; $\langle I_{z} \rangle$ becomes negative (positive) due to the negative (positive) $S_z$ injected optically by $\sigma^{+(-)}$ excitation. 
The $z$-component of $\bm B_{\rm n}$, ${B}_{{\rm n},z}$, is written by ${B}_{{\rm n},z}$=$\tilde{A} \langle {I_z} \rangle / {g}_{z}^{\rm e} \mu_{\rm B}$~\cite{OptOrientation} and the direction (sign) is determined by the signs of $\langle {I_z}  \rangle$, and $g_{z}^{\rm e}$ considering $\tilde{A}>0$ for III-V QDs. Although the sign of $\langle {I_z} \rangle$ depends on the excitation polarization, 
${B}_{{\rm n},z}$ is generated always antiparallel to $\langle I_{z} \rangle$ for ${g}_{z}^{\rm e}<0$ as shown in Fig.~\ref{Fig2}(a). Inversely, ${B}_{{\rm n},z}$ is generated parallel to $\langle I_{z} \rangle$ for ${g}_{z}^{\rm e}>0$ as shown in Fig.~\ref{Fig2}(b). 
Therefore, the compensation of positive $B_z$ with ${B}_{{\rm n},z}$ and NSSW occurs at $\sigma^{-(+)}$ excitation for ${g}_{z}^{\rm e}<0 \ ({g}_{z}^{\rm e}>0)$. Note that the compensation by ${B}_{{\rm n},z}$ affects only the electron spin states.

The above mentioned excitation polarization dependence of the electron states is reflected directly to the change of $\Delta E$ affected by $\bm B_{\rm n}$ as shown in the lower panels of Fig.~\ref{Fig2}. 
When ${B}_{{\rm n},z}$ is formed antiparallel to $B_z$, $\langle I_{z} \rangle$, thus ${B}_{{\rm n},z}$, indicates the abrupt change by NSSW due to the positive feedback of the $\langle I_{z} \rangle$ formation rate~\cite{Maletinsky07,Urbaszek13}, and a large change of $\Delta E$ emerges. However, the change of $|\Delta E|$ by NSSW is observed as an abrupt \textit{reduction} of $|\Delta E|$ for ${g}_{z}^{\rm e}<0$ (Fig.~\ref{Fig2}(a)) and an abrupt \textit{increase} of $|\Delta E|$ for ${g}_{z}^{\rm e}>0$ (Fig.~\ref{Fig2}(b)). 

The induced abrupt change of $|\Delta E|$ and the sign of $\Delta E$ are summarized in Table~\ref{TBL1}. 
In the cases of ${g}_{z}^{\rm h}>0$, the pattern of $\Delta E$ is shifted in negative region according to the definition of $\Delta E$. 
Note that the change of $|\Delta E|$ by NSSW is determined by the product ${g}_{z}^{\rm e}{g}_{z}^{\rm h}$~\cite{Kaji07} because the product determines that the pattern of the transitions with $\sigma^+$ and $\sigma^-$ polarizations as shown in the upper panels of Fig.~\ref{Fig2}(a) and (b); one is the included pattern (a) and the other is the nested pattern (b). 
In the figures of $\Delta E$ (lower panels), the Zeeman splitting of the hole spin states, ${\rm |g}_{z}^{\rm h}|\mu_{\rm B} B_z$, is indicated by a horizontal thin solid line. The variation from the thin line corresponds to the electron Zeeman splitting ${\rm |g}_{z}^{\rm e}|\mu_{\rm B} (B_z \pm B_{{\rm n},z})$.  
Thus, the Zeeman splitting of two PL lines by the linearly polarized excitation (i.e. without $B_{\rm n}$) is given by $(|g^{\rm h}_z|+ |g^{\rm e}_z|)\mu_{\rm B}B_z$ for the former pattern and $(|g^{\rm h}_z|- |g^{\rm e}_z|)\mu_{\rm B}B_z$ for the latter pattern. Considering the sign of the g factors, the g factors of $X^0$ and $X^+$ can be written by ${g}_{z}^{\rm h}+{g}_{z}^{\rm e}$. 
From Figs.~\ref{Fig1} and ~\ref{Fig2}, ${g}_{z}^{\rm e}>0$ and ${g}_{z}^{\rm h}<0$ (Fig.~\ref{Fig2}(b)) can be assigned for the observed single InAlAs QD. 
Although the magnitude $|{g}_{z}^{\rm e}|$ and $|{g}_{z}^{\rm h}|$ as well as the signs can be obtained from Fig.~\ref{Fig1}, the error of the magnitude has to be deduced from a more precise measurement by changing $B_z$ to a few values (not shown here). As a result, ${g}_{z}^{\rm e}$=$+0.34\pm$ 0.02 and ${g}_{z}^{\rm h}$=$-2.57\pm$0.01 are obtained for the studied single InAlAs QD. 
In addition, the magnitude of the in-plane g factor ($|{g}_{\perp}^{\rm e}| \sim |{g}_{\perp}^{\rm h}|$=0.35$\pm$0.01) and the in-plane anisotropy of the QD have been already reported~\cite{Yamamoto18}. From the isotropic nature of the conduction band, the ${g}_{\perp}^{\rm e}$ is considered to have a positive sign too.

\begin{table}[tb]
\caption{Sign identification of g factors based on the change and sign of $\Delta E$. }
\label{TBL1}
\begin{tabular}{c|cccc}
\hline
& \multirow{2}{*}{Sign of ${g}_{z}^{\rm e},\ {g}_{z}^{\rm h}$}  & change of $|\Delta E|$ & \multirow{2}{*}{sign of $\Delta E$}& \multirow{2}{*}{Fig.~\ref{Fig2}}\\
&  & by NSSW &  & \\
\hline
\multirow{2}{*}{${g}_{z}^{\rm e}\cdot{g}_{z}^{\rm h} > 0$} &${g}_{z}^{\rm e} > 0$, ${g}_{z}^{\rm h} > 0$ & reduction & negative &  \\ 
 &${g}_{z}^{\rm e} < 0$, ${g}_{z}^{\rm h} < 0$ & reduction & positive & (a) \\
 \hline
\multirow{2}{*}{${g}_{z}^{\rm e}\cdot{g}_{z}^{\rm h} < 0$}  &${g}_{z}^{\rm e} < 0$, ${g}_{z}^{\rm h} > 0$ & increase & negative & \\
 &${g}_{z}^{\rm e} > 0$, ${g}_{z}^{\rm h} < 0$ & increase& positive & (b)\\ 
\hline
\end{tabular}
\end{table}

\begin{figure}[tb]
  \begin{center}
    \includegraphics[width=165pt]{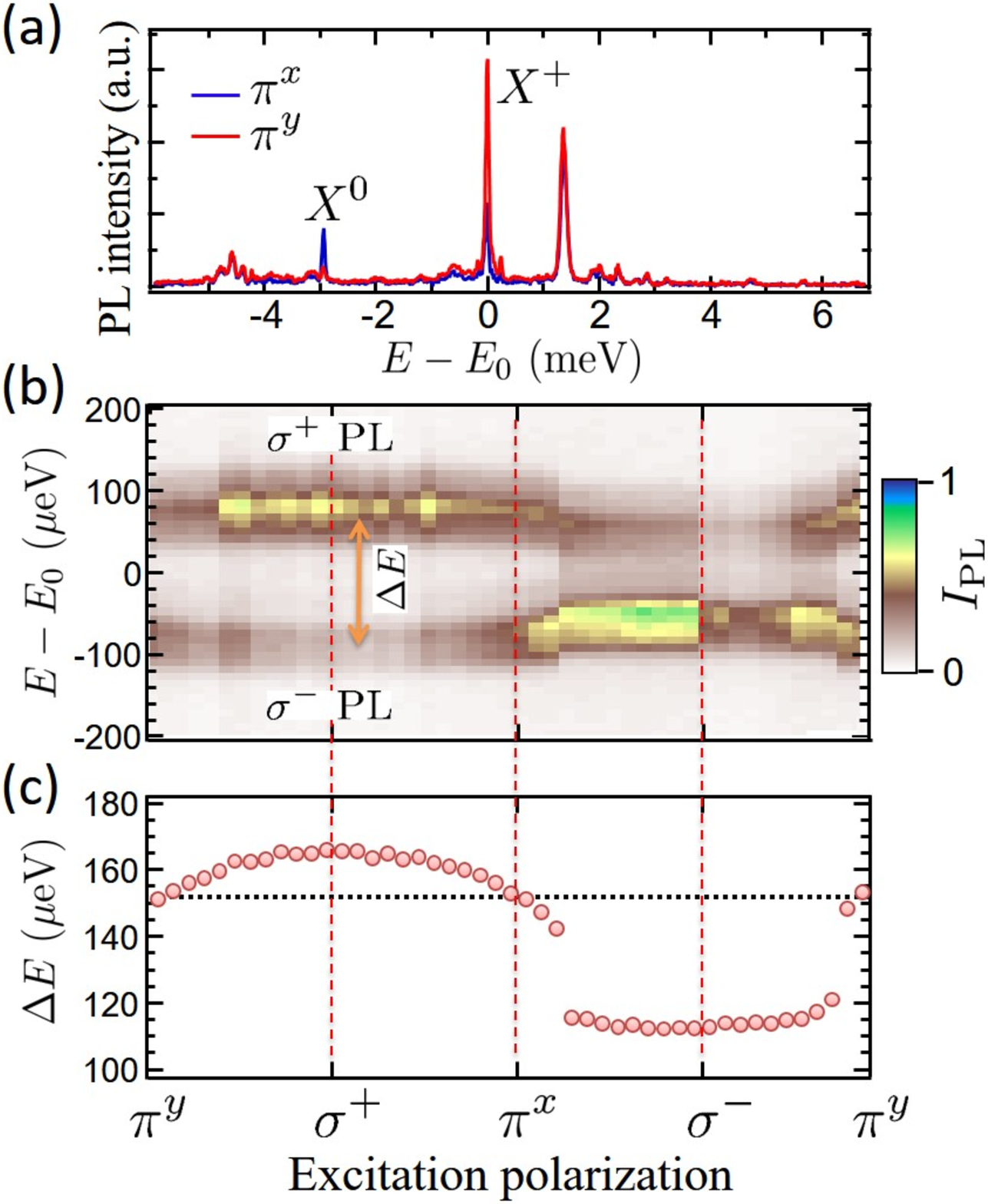}  
  \end{center}
  \caption{(color online) (a) Polarization-resolved  PL spectra of a typical single InAs/GaAs QR at 0 T under the nonpolarized excitation. The horizontal axis is replotted from the $X^{+}$ PL peak energy $E_0$=1.3330 eV. (b) A two-dimensional plot of the charged exciton emission in the single InAs/GaAs QR under 1 T. The excitation polarization was varied systematically from $\pi^x$ to $\sigma^{+}$ and $\sigma^{-}$. (c) The change in $\Delta E$ depending on the excitation polarization.}
  \label{Fig3}
\end{figure}

Next, the method is applied to identify the sign of g factors in single InAs/GaAs QRs. 
The averaged QR sizes are $\sim$40 nm in outer diameter, $\sim$10 nm in inner diameter, and $\sim$10 nm in height. 
The details of the growth conditions are seen in Ref.~\onlinecite{Tu_APL07}. The magnitude of out-of-plane ($|g_{z}^{\rm e}|$, $|g_{z}^{\rm h}|$) and in-plane g factors ($|{g}_{\perp}^{\rm e}|$, $|{g}_{\perp}^{\rm h}|$), and their anisotropy for many QRs are already reported in Refs.~\onlinecite{Kaji14} and~\onlinecite{Kaji_pssb17}. Under $B_{z}=+$1.0 T, the excitation polarization dependence measurements with the same setup as those in InAlAs QDs were carried out. The polarization-resolved PL spectra in Fig.~\ref{Fig3}(a) indicates very small FSS less than our spectral resolution, which is expected from the annealing in the process of QR formation from QD~\cite{Tu_APL07}.  
Figure~\ref{Fig3}(b) shows a two-dimensional plot of the excitation polarization dependence of the $X^+$ PL spectra. The corresponding energy splitting ($\Delta E$) is plotted in Fig.~\ref{Fig3}(c). Unlike the case of InAlAs QDs, an abrupt reduction in $\Delta E$ was observed around the $\sigma^{-}$ excitation, which corresponds to the response shown in Fig.~\ref{Fig2}(a), that is, ${g}_{z}^{\rm e} < 0$ and ${g}_{z}^{\rm h} < 0$. Further, we confirmed that many other InAs QRs grown in the same sample indicated the similar responses (not shown here). Consequently, ${g}^{\rm e}_z$=$-0.51 \pm$0.02, ${g}^{\rm h}_z$=$-2.10\pm$0.02 are obtained for this single InAs QR. 

Previously, the sign of g factors of InGaAs/GaAs QDs was examined experimentally from the Zeeman splitting of bright and dark excitons, which is one of clear methods to determine the sign of g factors although it is necessary to select the QD with a reduced shape symmetry or tilted quantization axis. For In$_{0.6}$Ga$_{0.4}$As/GaAs QDs, ${g}^{\rm e}_z$=$-0.81$ and ${g}^{\rm h}_z$=$-2.21$ were reported~\cite{Bayer99}, and the results were supported by the theoretical studies~\cite{Nakaoka04}. The sign of g factors in InAs QRs coincides with the results in In(Ga)As QDs.
From both measurements of Figs.~\ref{Fig1} and~\ref{Fig3}, it is found that the sign of $g_{z}^{\rm e}$ in InAlAs QDs is opposite to the one in InAs QRs while the signs of $g_{z}^{\rm h}$ in the both systems are the same.   

Finally, we confirm the conditions adopted in this work. 
From Figs.~\ref{Fig1} and~\ref{Fig3}, it is clear that the DCP of the PL emission varies following the change of the excitation polarization, that is the case of a positive DCP. Here, the DCP is defined by $(I^+ - I^-)/(I^+ + I^-)$ with the integrated intensity $I^{+(-)}$ of the  $\sigma^{+(-)}$ polarized PL spectrum. It means that the optically-excited electron spin polarization $\langle S_z \rangle$ was highly preserved right before the radiative recombination and contributes to form $B_{\rm n}$. If the negative DCP is observed for $X^+$ PL (i.e. the $\langle S_z \rangle$ is flipped before the recombination from the ground state), the signs of $g_z^{\rm e}$ and $g_z^{\rm h}$ are reversed respectively. 
In the case of $X^-$, the interacting electron spin that form $B_{\rm n}$ by HFI is the remained electron spin after the radiative recombination. Therefore, the reversed electron spin is effective for $B_{\rm n}$ formation in the positive DCP case. In contrast, when the hole spin in $X^-$ is flipped during the relaxation process to the ground state and the negative DCP is observed, the same spin as the optically-excited electron is effective for $B_{\rm n}$ formation. We summarize the application of the proposed method to two charge states in Table~\ref{TBL2}. 



\begin{table}[tb]
\caption{Application of the method to two charge states. The letter A (R) means that the signs of $g_z^{\rm e}$ and $g_z^{\rm h}$ in Table~\ref{TBL1} can be applied without change (reversely).}
\label{TBL2}
\begin{tabular}{ccc}
\hline
 &  \multicolumn{2}{c}{charge state of the ground level}\\
DCP &  $X^+$ & $X^-$ \\
\hline
positive   & A & R \\
negative   & R & A \\
\hline
\end{tabular}
\end{table}

Before summarizing this work, we should denote the limitation of this sign identification method. Our method can be applied to the PLs of any charge states originated from single nanostructures like QDs of commonly used III-V direct-gap semiconductors as long as NSSW occurs there. This is because the III-V elements of the well-established compound semiconductors have non-zero nuclear spins with large natural abundance. The minimal magnitude of the g factors is  $\sim$0.02 for evaluation in our current system, which depends basically on the spectral resolution ($\sim$5 $\mu$eV).

In summary, we demonstrated a convenient sign identification method of electron and hole out-of-plane g factors utilizing the correlation of NSSW and the sign of Zeeman splitting. 
The usefulness of the method was demonstrated for individual In$_{0.75}$Al$_{0.25}$As/Al$_{0.3}$Ga$_{0.7}$A QDs and InAs/GaAs QRs, where the sign of the electron g factor was different. 
Although we vary the excitation polarization systematically for the demonstration, the magnitude and sign of the out-of-plane g factors can be deduced by observation of the PL spectra at only three excitation polarization $\pi^y$ (or $\pi^x)$, $\sigma^-$, and $\sigma^+$ under a positive $B_z$. Further, this method offers advantages to evaluate the fluctuation of $B_{\rm n}$ directly by combining the electron Zeeman part of $\Delta E$ and DCP in the case of $X^+$~\cite{Matsusaki17}.

The authors would like to acknowledge H. Sasakura in Hokkaido University and S.-J. Cheng in National Chiao-Tung University for helpful discussions. This work is supported by JSPS KAKENHI (Grants No. JP26800162 and No. JP17K19046) and the Asahi Glass Foundation (Japan).




\begin{thebibliography}{9}
\bibitem{comment} Two definitions of the sign of hole g factor with the same Zeeman Hamiltonian has been used in the previous works; $g_z^{\rm h}>0$ is assigned for the situation that the up-spin hole state $| +3/2 \rangle$ has a higher energy and for the another situation that the up-spin valence electron state $| -3/2 \rangle$ has a higher energy.
\end{thebibliography}
\end{document}